%Paper: hep-ph/9303220
%From: COLANGELO@BARI.INFN.IT
%Date: Thu,  4 MAR 93 13:54 GMT

\hoffset=0.141in
\voffset=0.33in
\magnification=\magstep1
\hsize=146truemm
\vsize=222truemm
%\nopagenumbers
%
\baselineskip=14pt
\font\small=cmr9
\def\spc{\hskip 3 pt}
\def\slash#1{#1\hskip-0.5em /}
\hfill {\vbox {\hbox {BARI TH/93-132} \hbox {UTS-UFT-93-3}
\hbox {\it January 1993} }}
\vskip 30 pt
\centerline{ {\bf QCD SUM RULES AND B DECAYS}
\footnote{$^{(*)}$}{Contributed paper to the ECFA Workshop on the Desy B
Factory} }
\bigskip
\medskip
\centerline{Pietro Colangelo {\it $^{a}$},  Giuseppe Nardulli {\it $^{a,b}$}}
\smallskip
\centerline {\it $^{a}$Istituto Nazionale di Fisica Nucleare,
Sezione di Bari, Italy}
\smallskip
\centerline {\it $^{b}$Dipartimento di Fisica dell'Universit\`a di Bari, Italy}
\bigskip
\centerline {Nello Paver}
\smallskip
\centerline {\it Dipartimento di Fisica Teorica dell'Universit\`a di Trieste,
Italy}
\smallskip
\centerline {\it Istituto Nazionale di Fisica Nucleare, Sezione di Trieste,
Italy}
\vskip 60pt
\centerline {\bf Abstract}
\midinsert\narrower\narrower
\par\noindent We review some recent applications of QCD sum rules to
$B$-decays. These include the determination of the leptonic constant $f_B$
and of the semileptonic transition amplitudes of the $B$-meson into
negative and positive parity charmed states.
\endinsert
\vfill\eject
%\centerline { }
%\vskip 10cm
\noindent {\bf 1. Introduction}\hfill\break
\medskip
Considerable interest has been given in the past years to applications
of QCD sum rules to $B$ (and $D$) physics, resulting in a large number of
estimates for the $B$ and $D$ transition matrix elements.\par
The formalism consists in relating {\it via} dispersion relations
low energy parameters, such as hadronic masses and couplings, to the
QCD dominated short-distance Operator Product Expansion (OPE) of current
correlators [1,2]. Such an expansion is assumed to hold in the
presence of
non-perturbative QCD effects, which are parametrized by a set of
fundamental vacuum condensates, describing the breaking of asymptotic
freedom, and inducing power corrections in the
correlators. These condensates are fundamental constants which
cannot be computed from perturbation theory: however, they are ``universal''
and, once fitted from channels where the physics is known,
their values can be
used to make predictions for the other channels.
Thus, being fully field-theoretic and relativistic, and incorporating the
fundamental features of QCD, this method aims to be a means
 to interface from first
principles hadronic observables to the basic quark and gluon QCD
Lagrangian. From this point of view, it should represent a convenient
framework to study the properties of heavy-light quark systems and of
quarkonia, which avoids the notion of constituent quark bound state
wavefunction and should thus reduce the model dependence to a
minimum.\par
Most recently, the attention has been focussed on the applications of QCD
sum rules to the semileptonic and the purely leptonic $B$- and $D$-meson
decays, motivated on one side by the observed features
of the data [3,4] (also in the perspectives of increased
precision and of rare decays measurements,
as allowed by planned charm and beauty factories), and on the
other side by the theoretical achievements offered in principle by
the heavy quark effective theory (HQET) [5]. In this regard, it turns
out that QCD sum rules allow to study the behaviour of
amplitudes for $\bar qQ$ mesons for ${m_Q\rightarrow\infty}$ ($Q$
represents the heavy quark, $q$ the light one).
Therefore they can be used to
verify the predictions
of spin-flavor heavy quark symmetry, and to  test quantitatively
the resulting scaling laws relating the different amplitudes as well as
the validity of approximations based on the $1/m_Q$ expansions.
Accordingly, the sequel will be devoted to applications of QCD sum rules
to several  aspects of $B$ physics. Specifically,
we shall briefly review the predictions for the leptonic constant $f_B$,
analyzing in particular the determination of the beauty quark mass $m_b$,
and of the
semileptonic $B$-decays to negative and to positive parity charmed mesons.
\bigskip
\noindent{\bf 2. The leptonic decay constant $f_B$}
\medskip
Leptonic decay constants, governing the decays $D\rightarrow\mu\nu$,
$B\rightarrow\mu\nu$, {\it etc.}, are defined as the vacuum-to-meson
matrix elements:
$$<0|{\overline Q}\gamma_{\mu}\gamma_5q|P(p)>=if_P\ p_{\mu},
\eqno(1)$$
with $Q=c,b$ the heavy quarks and $q=u,d,s$ the light ones, so that $P=B,D$.
The theoretical interest stems from their significant role as
normalizations of numerous predictions for heavy meson transitions.
In particular, by dominating the description of $B-\bar B$ mixing, $f_B$
strongly affects the determination of the CKM matrix, with important
implications on CP violation in the beauty sector [6]. Furthermore,
of great interest is the scaling behaviour for very large (infinite) $m_Q$:
$$f_P={const\over{\sqrt{m_P}}},\eqno(2)$$
modulo logs, which reflects the connection with the constituent quark
wavefunction, but in fact results from the heavy quark symmetry and
accordingly is a
general consequence of QCD.\par
The present indications are $f_D< 310 \spc MeV $ [7] and
$f_{D_s}=(232\pm45\pm20\pm48)$ $MeV$ [8], whereas no information is
available on $f_B$. To get a feeling on the sensitivity required to
measure $f_B$, for $|V_{ub}/V_{cb}|=0.1$ with $|V_{cb}|=0.045$
and $f_B=1.5f_{\pi}$ (${f_{\pi}=132\ MeV}$)
one finds $BR(B\to\mu\nu)=5.1 \times 10^{-7}$ and
$BR(B\to\tau\nu)=1.1 \times 10^{-4}$.\par
Let us now briefly describe  the method of QCD sum rules
as applied to $f_B$. One  starts from
the two-point correlator at euclidean $Q^2=-q^2$:
$$\Pi_5(q^2)=i\int d^4x \exp (iqx)
<0|T\left(J_5(x)J_5(0)^{\dagger}\right)|0>,\eqno(3)$$
where $J_5(x)=:{\overline Q}(x)i\gamma_5 q(x):$ is the quark bilinear
operator which interpolates the pseudoscalar meson field, and one can
neglect the light quark mass $m_q$ with respect to $m_Q$. For
other mesonic states, the appropriate interpolating fields must be used
in (3), {\it e.g.} $V_{\mu}=:{\overline Q}\gamma_{\mu}q:$ for the vector
particle $B^*$, {\it etc}. In QCD $\Pi_5(q^2)$ satisfies a twice-subtracted
dispersion relation. Actually, to improve the convergence, emphasize
the contribution of the hadronic ground state one is interested in, and get
rid of unknown subtraction constants, suitable weight functions are used in
the dispersive integrals. One is thus led to consider ``moments'', each
corresponding to a different version of QCD sum rules,
such as the Hilbert moments
at $Q^2=0$, leading, for $n=1,2\dots$, to the finite energy sum rules
(FESR):
$$M_n(0)\equiv {{(-1)^n}\over{(n+1)!}}\left({d\over{dQ^2}}\right)^{n+1}
\Pi_5(Q^2)\big|_{Q^2=0}=\int_0^\infty ds{1\over{s^{n+2}}}{1\over{\pi}}
Im \Pi_5(s);\eqno(4)$$
and the exponential moments
$${\widetilde\Pi}_5(\sigma)=\int_0^\infty
ds \exp{(-\sigma s)}{1\over{\pi}}Im \Pi_5(s),\eqno(5)$$
where $1 / \sqrt \sigma$ is a mass scale. A review of other versions
can be found in [2].\par
The hadronic spectral function on the right sides of (4) and (5) is usually
parametrized in terms of the ground state meson pole plus a ``continuum''
of higher
states, starting at some threshold $s_0$ and modelled by the asymptotic
freedom (AF) expression:
$${1\over{\pi}}Im\Pi_5(s)\big|_{HAD}={f_P^2m_P^4\over m_Q^2}\delta(s-m_P^2)
+{1\over{\pi}}Im\Pi_5(s)\big|_{AF}\theta(s-s_0).\eqno(6)$$
\noindent In (6), to two loops [9]:
$$\eqalignno{{1\over{\pi}}Im\Pi_5(s)\big|_{AF}&={3\over{8\pi^2}}
m_Q^2{{(1-x)^2}\over x} \biggl\{ 1+{{4\alpha_S}\over{3\pi}}\biggl[{18\over 8}
-2Li_2\left({{-x}\over{1-x}}\right)\cr
&-\ln\left({x\over{1-x}}\right)\ln\left({1\over{1-x}}\right)+\left({3\over 2}-
{x\over{1-x}}-x\right)\ln\left({x\over{1-x}}\right)\cr
&+{1\over{1-x}}\ln\left( {1\over{1-x}}\right)\biggr]\biggr\} ,&(7)\cr}$$
where $x\equiv m_Q^2/s$.\par
The left sides of (4) and (5) can be evaluated by means of the
OPE, in terms of short-distance perturbative QCD,
determined by the integral of (7) from $m_Q^2$ up to infinity, plus
non-perturbative power corrections,
determined by quark and gluon operator vacuum condensates, ordered for
increasing dimension. These vacuum expectation values essentially allow the
extrapolation of the asymptotic freedom contribution, completely known in
terms of $\alpha_s$ and  $m_Q$, down to ``moderate'' mass scales, relevant
to hadrons.\par
Limiting to operators of dimensionality up to six, combining (4), (6) and (7)
one obtains the FESR, in the form ($m_q\rightarrow 0$):
$${f_P^2\over m_P^{2n}}=m_Q^2
\int\limits_{m_Q^2}^{s_0}{ds\over s^{n+2}}{1\over{\pi}}Im\Pi_5\big|_{AF}
+M_n(0)\big|_{NP},\eqno(8)$$
where the non-perturbative part is given by:
$$\eqalignno{M_n(0)\big|_{NP}&={-m_Q\langle\bar q q\rangle\over m_Q^{2n+2}}
\left[1-{\langle\alpha_s G^2\rangle\over 12\pi m_Q\langle\bar q q\rangle}
-{1\over 4}(n+2)(n+1){M_0^2\over m_Q^2}\right.\cr
&\left.-{4\over 81}(n+2)(n^2+10n+9)\pi\alpha_s\rho{\langle\bar q q\rangle\over
m_Q^3}\right].&(9)\cr}$$
Alternatively, combining (5), (6) and (7) one obtains the
exponential (also called  Lapla\-ce or Borel) version of QCD sum rules
for $f_P$ [10]:
$${f_P^2m_P^4\over m_Q^2}\exp{\left(-\sigma(m_P^2-m_Q^2)\right)}=
\int\limits_{m_Q^2}^{s_0}ds\exp{\left(-\sigma(s-m_Q^2)\right)}
{1\over{\pi}}Im\Pi_5\big|_{AF}+{\widetilde\Pi}_5(\sigma)\big|_{NP}
\eqno(10)$$
where
$$\eqalignno{{\widetilde\Pi}_5(\sigma)\big|_{NP}&=
-m_Q\langle\bar q q\rangle\left[1-
{\langle\alpha_s G^2\rangle\over 12\pi m_Q\langle\bar q q\rangle}
+{M_0^2\sigma\over 2}\left(1-{m_Q^2\sigma\over 2}\right)\right.\cr
&\left.+{8\sigma\over 27}{\pi\alpha_s\rho\langle\bar q q\rangle\over
m_Q}\left(2-{m_Q^2\sigma\over 2}-{m_Q^4\sigma^2\over 6}\right)\right].
&(11)\cr}$$
Sum rules for the meson mass $m_P$ can be obtained from ratios of FESR with
different $n$, or from the logarithmic derivative of (10) with respect to
$\sigma$. Applications of other versions of QCD sum rules to the estimate of
$f_P$ are presented {\it e.g.} in [11] and [2].\par
In (9) and (11), $M_0^2$ parametrizes the dimension 5 quark-gluon condensate
${\langle g_s\bar q\sigma_{\mu\nu}G^{\mu\nu}q\rangle=
M_0^2\langle\bar qq\rangle}$, and $\rho \ne 1$ represents the deviation of the
dimension 6 four-quark condensate from pure factorization. Typical values,
obtained from PCAC and from applications of the method to other channels,
are {\it e.g.}
${\displaystyle\langle\bar q q\rangle(1\ GeV)\simeq -0.016\ GeV^3}$,
$\langle\alpha_s G^2\rangle\simeq 0.02\div 0.06\ GeV^4$, $M_0^2\simeq
0.5\div0.8\ GeV^2$ and $\rho\simeq 1\div 3$. In practice,
results for $f_B$ are not
sensitive to the value of the gluon condensate and $\rho$. Conversely,
more important is their dependence on the input values of quark
masses, typical values being $m_b(Q^2=m_b^2)\simeq 4.6\div 4.8\ GeV$ and
$m_c(Q^2=m_c^2)\simeq 1.3\div 1.4\ GeV$.\par
Regarding the practical use of QCD sum rules, we remark that the
non-perturbative expansions in (9) and (11) have been truncated to the first
few significant terms (of lowest dimension), neglecting higher condensates.
The validity of such an approximation clearly depends on the values of $n$
or $\sigma$. Indeed, larger $n$ and $\sigma$ emphasize the ground state
in (8) and (10) and minimize the role of the continuum, but
correspondingly also increase the size of higher power corrections.
Thus, in practical applications one considers as a
compromise the first few values of $n$, or a suitable range of $\sigma$,
such that the truncation of the OPE is justified.
Another point concerns the role of the continuum threshold $s_0$, which is
not determined {\it a priori}.
Clearly, the predictions for the hadronic observables
must not be sensitive to the values of this parameter. Thus, the procedure to
exploit the FESR (8) is to search for a ``duality window'',
{\it i.e.} a range in which the predicted $m_B$ (and $m_D$) agrees with the
measured one and is stable against $s_0$ (and $n$). In this window one then
solves for $f_B$, and this estimate should be reasonably reliable.
Similarly,
in the case of the exponential sum rules (10) one imposes stability of
the predicted $m_P$ and $f_P$ in the range of $\sigma$ and
$s_0$ mentioned above.\par
In Table 1 we list the results for $f_B$ (and for completeness also for
$f_D$) previously obtained from FESR and from exponential
sum rules [12,13,14], using $\alpha_s$ and
$\langle\bar q q\rangle$ evaluated at a scale $m_Q$.
The authors of [10] use $m_b=4.8 \spc GeV$,
which is somewhat larger than the values of $m_b$ used by the other groups.
This shows that
the method is sensitive to the input value of the heavy quark mass.
The ``$\pm$'' in Table 1 accounts for reasonable variations of the
input parameters. Within $SU(3)$ breaking effects, the results for
$f_D \simeq f_{D_s}$ are compatible with the experimental data [7].
\def\boxit#1{\vbox{\hrule\hbox{\vrule\kern3pt\vbox{\kern3pt#1\kern3pt}
\kern3pt\vrule}\hrule}}
\setbox9=\vbox{\hsize 25.pc \noindent \strut
{$$\vbox{\settabs
\+ Laplace and Hilbert & $1.38 \pm 0.14$ \hskip 10pt
& $1.31 \pm 0.12$ \hskip 10pt & [13] \cr
\+ Method    & $f_B/f_\pi$   & $f_D/f_\pi$  & Ref. \cr
\+    & & & \cr
\+ Laplace & $1.02 \pm 0.11$ & $1.33 \pm 0.19$ & [10] \cr
\+ Hilbert moments & $1.35 \pm 0.25$ & $1.70 \pm 0.20$ & [12] \cr
\+ Laplace \& Hilbert & $1.38 \pm 0.14$ & $1.31 \pm 0.12$ & [13] \cr
\+ Laplace & $1.29 \pm 0.15$ &  &  [14] \cr
\+ Laplace (HQET) & $1.48 \pm 0.39$ & $1.29 \pm 0.23$ & [17] \cr
\+ Laplace (HQET) & $1.7 \pm 0.2$ &   & [19] \cr
}$$ } }

$$\boxit{\box9}$$
\centerline {\small {\bf Table 1} QCD sum rules predictions for $f_P$}
\vskip 1cm
\par
Another important point is that the size of the
$O(\alpha_s)$ corrections in the
sum rule for $f_B^2$ turn out to be rather large, of the order of 40\%.
Moreover, at this two loop level
there is an ambiguity in the definition of the
argument of $\alpha_s$,
{\it i.e.} whether it should be $m_b$ or the much smaller confinement scale
of the order of $1\ GeV$, which can make a substantial numerical difference.
\par
Indeed, in this regard it is instructive to consider the ``static
approximation'' of (10), where $m_Q$ becomes infinitely heavy.
This can formally be obtained by defining the ``non-relativistic'' variables:
$$\sqrt s=m_Q+\omega;\qquad m_P=m_Q+\omega_R;\qquad \tau=2m_Q\sigma,\eqno(12)$$
where $\tau$ is a euclidean time and the ``binding energy'' $\omega_R$ should
be flavor-independent in the heavy quark limit. Expanding up to the first
order $1/m_Q$, and retaining the condensates up to dimension 5,
Eq.(10) becomes [15,16,17]:
%\vfill\eject
$$\eqalignno{&\left(f_P\sqrt{m_P}\right)^2\left({m_P\over m_Q}\right)^3
\exp{(-\tau\omega_R)}=\cr
&{3\over\pi^2}\int\limits_{0}^{\omega_0}d\omega\omega^2
\left(1-{2\omega\over m_Q}\right)\exp{(-\tau\omega)}
\biggl\{ 1+{\alpha_s\over\pi}\left[
2\ln{\left({m_Q\over 2\omega}\right)}+{13\over 3}+{4\pi^2\over 9}
-{4\omega\over m_Q}\right]\biggr\} \cr
&-\langle\bar q q\rangle\left[1-
{M_0^2\tau^2\over 16}\left(1-{4\tau\over m_Q}\right)
+{\langle\alpha_s G^2\rangle\over 12\pi m_Q\langle\bar q q\rangle}\right].
&(13)\cr}$$
Analogously to the relativistic version (10), a sum rule for $\omega_R$
can be obtained from the logarithmic derivative of Eq.(13) with respect
to $\tau$.\par
By construction, this non-relativistic form should reproduce the large $m_Q$
dimensional behaviour implied by (2), plus $1/m_Q$ corrections. In addition,
it should be noticed that, for $m_Q\rightarrow\infty$, the
right hand side of (13) has no explicit reference
to $m_Q$, and actually now involves mass scales of the order of the
confinement scale ${\sim 1\ GeV}$ (the same is expectedly true
for $\omega_R$). This fact naively suggests to take a low energy,
confinement scale as the argument of the running $\alpha_s$ and
$\langle\bar q q\rangle$.\footnote{*}
{\small\baselineskip 12pt
This suggestion was also made in
[18], discussing the application of QCD sum rules to quarkonium
states.} However, Eq.(13) clearly shows that there is a difficulty with the
$O(\alpha_s)$ correction, associated with the appearance of a large
(diverging) $\ln{m_Q}$, so that the limit $m_Q\rightarrow\infty$ is only
defined after a resummation of these terms to all orders.
Also, one can see that the finite coefficient in the $\alpha_s$ correction
is of the order of 10, giving rise to a large contribution, regardless of the
choice of the argument of the strong coupling constant. The separation of
the $\ln{m_Q}$ dependence from the sum rule and the definition of the
appropriate scale require the renormalization group improved summations of
the leading $(\alpha_s\ln{m_Q})^n$ and the next-to-leading
$\alpha_s(\alpha_s\ln{m_Q})^n$ terms to all orders in perturbation theory.
This has been recently done by formulating the QCD sum rules in the framework
of the HQET [17,19].\par
Basically, in this theory one has to investigate the two-point correlator
$$\Pi_5(w)=i\int d^4x \exp (ikx)
<0|T\left({\tilde J}_5(x){\tilde J}_5(0)^{\dagger}\right)|0>,\eqno(14)$$
where ${\tilde J}_5=\bar q\gamma_5 h_Q(v)$ is the effective pseudoscalar
current, with $h_Q(v)$ the velocity-dependent heavy quark field. The ground
state contribution is essentially determined by
${\langle 0 \vert {\tilde J}_5\vert P(v)\rangle=\sqrt{m_Q}F(\mu)}$,
where $F(\mu)$ is an
effective theory low energy constant, independent of $m_Q$ but depending on a
renormalization scale $\mu$. The relation between the physical and the
HEQT $f_P$  is:
$$f_P\sqrt{m_P}=C_F({m_Q\over\mu})F(\mu)+O({\Lambda_{QCD}\over
m_Q}),\eqno(15)$$
where $C_F$ is a QCD short-distance coefficient, which is computed
up to the next-to-leading order [20,21,17]. The left side of (15)
must be $\mu$-independent; to the considered  order the $m_Q$- and the
$\mu$-dependences in ${\displaystyle C_F({m_Q\over\mu})}$ factorize, so that
(15) can be suitably rewritten as :
$$f_P\sqrt{m_P}={\hat C}_F(m_Q){\hat F},\eqno(16)$$
where
$${\hat C}_F(m_Q)=\left(\alpha_s(m_Q)\right)^{d/2}
\left[1+{\alpha_s(m_Q)\over\pi}Z\right],\eqno(17_a)$$
$${\hat F}=\left(\alpha_s(\mu)\right)^{-d/2}
\left[1-{\alpha_s(\mu)\over\pi}(Z+\delta)\right] F(\mu) \spc.\eqno(17_b)$$
In (17) $d=\gamma_0/\beta_0$ with $\gamma_0$ and $\beta_0$ the leading order
anomalous dimension and QCD $\beta$-function, $Z$ is a
scheme-independent constant and $\delta$ is scheme-dependent (for the
explicit
values we refer {\it e.g.} to [17]). According to ($17_b$), ${\hat F}$ is a
scale-independent, universal constant of QCD.\par
By the same procedure leading to (13) one obtains a QCD sum rule of the form
($m_Q=\infty$):
$$\eqalignno{F^2(\mu)&\exp{(-\tau\omega_R)}=
{3\over\pi^2}\int\limits_{0}^{\omega_0}d\omega\ \omega^2\exp{(-\tau\omega)}
\biggl\{ 1+{\alpha\over\pi} \left[2\ln{({\mu\tau\over 2})}+{13\over 3}\right.
\cr
&\left. +{4\pi^2\over 9}-2\ln{(\tau\omega)}+\delta \right] \biggr\}
-\langle\bar q q\rangle(\mu)\left[1-{M_0^2\tau^2\over 16}+\cdots\right],
&(18)\cr}$$
where the ellipses denote contributions of order $\alpha_s$ and of higher
dimensional condensates. Eq.(18) corresponds to (13) at $m_Q=\infty$,
except from the replacement of $m_Q$ by $\mu$ in the {\it log} and the
appearance of $\delta$. Both the $\mu$-dependence and $\delta$ must cancel
when considering the sum rule for the renormalization-group invariant
${\hat F}$. Indeed, the summation of the $\ln{\mu}$ to all orders
produces the factor ${\displaystyle\left(\alpha_s(2/\tau)/\alpha_s(\mu)
\right)^{-d}}$ and, taking into
 account ($17_b$), the running $\alpha_s$ must be
evaluated at a low-energy scale of the effective theory. The structure of the
QCD sum rule for ${\hat F}$ is finally [17]:
%\vfill\eject
$$\eqalignno{\left(\alpha_s(2/\tau)\right)^{d}&
\left(1+2{\alpha_s\over\pi}Z\right){\hat F}^2\exp{(-\tau\omega_R)}\cr
&={3\over\pi^2}\int\limits_{0}^{\omega_0}d\omega\ \omega^2
\exp{(-\tau\omega)}\left[1+{\alpha_s\over\pi}\left({13\over 3}
+{4\pi^2\over 9}-2\ln{(\tau\omega)}\right)\right]\cr
&-\langle\bar q q\rangle\vert_{2/\tau}
\left[1-{M_0^2\tau^2\over 16}+\cdots\right],
&(19)\cr}$$
where $\alpha_s$ is evaluated at a confinement scale.\par
The analysis then proceeds by looking for ranges in $\tau$ and in
$\omega_0$ where the prediction for ${\hat F}$ is stable. The results
exhibit a dependence on the value of $\omega_R$, which is typically of the
order of $0.5\div 0.8\ GeV$, depending on the value of the heavy
quark mass $m_b$ (or $m_c$). Such a dependence somehow limits the accuracy of
the predictions. Also, there is an uncertainty associated to
the actual choice for the low energy scale in the running strong coupling
constant. Although the different choices formally have an impact at the
$\alpha_s^2$ level, in practice they make a substantial difference because
the $\alpha_s$ corrections are large by themselves.\par
Combining the various uncertainties,
the final result for the ``static limit'' of $f_B$, defined as
$$f_B^{stat}={{\hat C}(m_b){\hat F}\over\sqrt{m_B}},\eqno(20)$$
is ${\displaystyle f_B^{stat}=200\div 300\ MeV}$, in good agreement with
the value recently found from lattice calculations [22]. However,
the price is represented by a 100\% order $\alpha_s$ correction, which
questions the validity of the perturbative expansion in this case.\par
Concerning the $O(1/m_Q)$ corrections due to the breaking of the heavy
quark symmetry, a systematic analysis is possible by combining (19) with
(13), and reintroducing the terms which vanish in the infinite mass limit
[17,19]. One source of corrections, which can be immediately assessed,
is related to the factor ${\displaystyle\left(m_P/m_Q\right)^3}$ on the
left hand side of (13). This shows that, although naively of order
${\displaystyle\left(\Lambda_{QCD}/m_Q\right)}$, the expansion parameter
is actually a factor {\it times}
${\displaystyle\left(\omega_R/m_Q\right)}$. For current values of quark
masses, $1/m_Q$ corrections are then expected to be reasonably small for
$f_B$, but very large for $f_D$.\par
The results of the complete analysis of the sum rules including $1/m_Q$
corrections are reported in Table 1, and one can compare them with the
values previously found from their relativistic counterparts. Within the
uncertainties, the overall picture indicates a leptonic constant of the $B$
comparable (or even larger) to that of the $D$, and in the range of possible
measurements at a beauty factory.
\bigskip
\noindent{\bf 3. A determination of the beauty quark mass}
\medskip
In the previous section we mentioned the dependence of the
QCD sum rule estimates of $f_B$ on the value of $m_b$. This is true also
for other transition matrix elements.
Thus, a precise determination of $m_b$ is highly desirable in order to reduce
the uncertainty in the theoretical predictions. The ideal source of
information on $m_b$ should be represented by the application of QCD sum rules
to the $\Upsilon$ system, where the data are extremely precise.
A typical result of such analysis, using exponential moments in the static
approximation, is $m_b = 4.65 \pm 0.05 \spc GeV$ [23].
It turns out that
estimates of $m_b$ are affected by
the approximations used to confront theory and experiment, and in particular
by the role of the $\alpha_s$ corrections [2,23].\par
In [18] the convenience of studying ratios of the exponential
moments (5) was pointed out. In fact, in ratios the
importance of higher order radiative corrections could be reduced considerably,
in particular in the non-relativistic limit. The latter corresponds to
infinite quark mass, and only the leading mass term was considered in
[18]. Moreover, since the ratios are found to depend sensitively on the
quark mass, they should be suitable for the extraction of $m_b$. As an
attempt to pursue improvements over the past analyses, it should be
interesting to reconsider the QCD sum rules proposed in [18],
including the next-to-leading mass corrections to the
non-relativistic limit [24].\par
To this aim, following the familiar procedure, we start from the correlator:
$$\eqalignno{\Pi_{\mu\nu}(q)&=
i\int d^4x \exp (iqx) <0|T\left(V_{\mu}(x)V_{\nu}(0)^{\dagger}\right)|0>\cr
&=\left(-g_{\mu\nu}q^2+q_{\mu}q_{\nu}\right)\Pi(q^2),&(21)\cr}$$
with $V_{\mu}(x)=\bar b(x)\gamma_{\mu}b(x)$. From the exponential moment
$$\Pi(\sigma)={1\over\pi}\int_0^{\infty} ds\exp{(-\sigma s)}Im\Pi(s),
\eqno(22)$$
we define the ratio, sensitive to the mass of the ground state:
$$R(\sigma)=-{d\over d\sigma}\left[\ln{\Pi(\sigma)}\right]={\Pi^\prime(\sigma)
\over\Pi(\sigma)},\eqno(23)$$
which tends to $m^2_{\Upsilon}$ for $\sigma\rightarrow\infty$.\par
In the non-relativistic limit, $\sqrt s=2m_b+\omega$, Eqs.(22) and (23)
become respectively:
$$\Pi(\tau)={1\over\pi}\int_0^{\infty}d\omega\exp{(-\tau\omega)}Im\Pi(\omega),
\eqno(24)$$
where $\sigma=\tau/4m_b$, and
$$R(\sigma=\tau/4m_b)=2m_b-{d\over d\tau}[ln{\Pi(\tau)}].\eqno(25)$$
At the two-loop level in perturbative QCD [25]:
$$\eqalignno{&{1\over\pi}Im\Pi(q^2)_{AF}=
	{1\over 8\pi^2}v(3-v^2)				\cr
	&\times\left\{1+{4\alpha_s\over 3}\left[{\pi\over 2v}-{(v+3)\over 4}
\left({\pi\over 2}
		-{3\over 4\pi}\right)\right]\right\}\times\Theta(s-4m_b^2),
	&(26)\cr}$$
where $v=\sqrt{1-4m_b^2/s}$ ($m_b$ is here the $b$-quark pole mass).
The leading non-perturbative contribution is given by the gluon condensate
[1]:
$$\Pi(q^2)_{NP}={1\over 48q^4}
\left[{3(v^2+1)(1-v^2)^2\over 2v^5}\ln{{1+v\over 1-v}}
-{3v^4-2v^2+3\over v^4}\right]\langle{\alpha_s\over\pi}G^2\rangle.\eqno(27)$$
Expanding (26) to leading plus next-to leading orders in
$m_b\rightarrow\infty$, we finally obtain the following QCD expression for
$R(\tau)$ defined in Eq.(25):
$$\eqalignno{R(\tau)&=2m_b\left\{1+{3\over 4m_b\tau}
\left(1-{5\over 6m_b\tau}\right)-
{1\over 3}\left({\pi\over m_b\tau}\right)^{1/2}\alpha_s(\tau)\right.\cr
&\times\left.\left[1-{2\over\pi m_b\tau}\left({19\over 2}\pi+{3\over16\pi}
-{5\over 8}\right)\right]+{\pi^2\over 48}\left({\tau\over m_b}\right)^2
\langle{\alpha_s\over\pi}G^2\rangle\right\},&(28)\cr}$$
where
$$\alpha_s(\tau)={12\pi\over 23\ln{(4m_b/\tau\Lambda^2)}}.\eqno(29)$$
The ratio (28) must be compared with a corresponding ratio involving the
experimental data. We parametrize the latter by the sum of narrow
$\Upsilon$ resonances followed by a hadronic continuum starting at
a threshold $s_0$, modelled by perturbative QCD as given in (26):
$$\eqalignno{\Pi(\sigma)_{EXP}=&{27\over 4\pi}{1\over\alpha_{EM}^2}\sum_{V}
\Gamma_V^{ee}m_V\exp{(-\sigma m_V^2)}\cr
&+{1\over\pi}\int_{s_0}^{\infty}
ds\exp{(-\sigma s)}Im\Pi(s)_{QCD}.&(30)\cr}$$
\par Analogously to (13), one can notice from (28) that in practice the
sum rule determines the difference between the hadronic ground state mass
$m_{\Upsilon}$ and $2m_b$, which is related to scales relevant to the
confinement dynamics.\par
The values of $m_b$ are then determined by matching the theoretical expression
(28) to the square root of the experimental ratio
$$R(\sigma)_{EXP}=-{d\over d\sigma}\ln{\Pi(\sigma)}_{EXP}.\eqno(31)$$
Stability requires the matching to occur in a range of $\sigma$ and $s_0$
(actually the sum rule is extremely well saturated by the $\Upsilon$ family,
so that the continuum has very little role), with corrections terms in
(28) at a safe, few percent level. Varying
$\Lambda_{QCD}$ and of $\langle\alpha_s G^2\rangle$
in the currently allowed range one finds:
$m_b=4.71\pm 0.05\ GeV.$
\par As a comment to this determination, we notice that
the subleading quark mass
cor\-rec\-ti\-ons can be important. For instance,
the $O(\alpha_s/m_b\sqrt{m_b})$ term in (28) is of the same size and
sign as the non-perturbative term, for the relevant values of $\tau$. Hence,
it would not be fully justified to keep the latter and ignore the former.
Thus, at this level, with all correction terms safely small, the
accuracy on $m_b$ is essentially limited by the uncertainties in
$\Lambda_{QCD}$, known to a factor of 2, and in the gluon condensate,
known to a factor of 3.
\bigskip
\noindent
{\bf 4. Semileptonic B decays into charmed final states in the limit
of infinitely heavy quarks}
\medskip
The hadronic matrix elements that describe the decays into negative-parity
states:
$$B \to D  \ell \nu\eqno(32)$$
\par\noindent and
$$B \to D^* \ell \nu\eqno(33)$$
\par\noindent
can be written in terms of form factors as follows:
$$<D (p')|\spc {\bar c}\gamma_\mu b|\spc B(p)> =
\spc G_+ \spc (p+p')_\mu + G_- \spc q_\mu\eqno(34)$$
$${1 \over i} \hskip 3pt <D^{*}(p',\epsilon)|\spc {\bar c}\gamma_\mu b
|\spc B(p)> =
\spc i \spc F \spc \epsilon_{\mu\nu\alpha\beta} \epsilon ^{*\nu}
(p+p')^\alpha q^\beta \eqno(35)$$
$${1 \over i} \hskip 3pt <D^{*}(p',\epsilon)|\spc {\bar c}\gamma_\mu
\gamma_5 b|\spc B(p)> =
\spc F_0 \spc \epsilon^*_\mu + F_+ \spc (\epsilon^* \cdot p) (p+p')_\mu +
F_- \spc (\epsilon^* \cdot p) q_\mu \eqno(36)$$
\par\noindent
where $G_\pm$, $F$, $F_{0,\pm}$ are functions of $q^2=(p-p')^2$.
\par
These form factors will be discussed for finite values of the heavy quark
masses in Section 6. Here we consider the results obtained in the limit
$m_Q \to \infty$. In the infinite
heavy quark
mass limit one can build up an effective field theory (HQET) [5]
and obtain useful predictions for the above form factors. More precisely,
one assumes that
$m_Q$ is much larger than the average momentum of the
light degrees of freedom $q^\mu$: $m_Q>>|q^\mu| .$
If  we write
$p_Q^\mu \spc =\spc m_Qv^\mu+k^\mu=m_Qv_Q^\mu$,
where $v^\mu$ is the hadron velocity, $v_Q^\mu$ is the quark velocity
and $k^\mu$ is a residual small momentum, in the limit
$$m_Q \to \infty \spc , \spc \spc \spc p_Q^\mu/m_Q \spc \spc  fixed \eqno(37)$$
we get
$$v_Q^\mu = v^\mu\eqno(38)$$
i.e. the heavy quark velocity is equal to the hadron velocity:
QCD interactions do
not change $v_Q^\mu$ that is therefore conserved. Eq.(38) is called
{\it velocity superselection rule}${}$ [5].

On can show that in the limit (37) the heavy quark propagator becomes:

$${\slash v +1 \over 2} {i \over vk}\spc ,\eqno(39)$$
\par\noindent
and the heavy quark gluon vertex assumes the form:

$$-i g T^a v_\mu \spc , \eqno(40)$$
\par\noindent
because, as a consequence
of the equations of the motion, the effective heavy quark field operator
$\displaystyle h_v(x) = {\slash v +1 \over 2}  e^{im_Q vx} Q(x)$
satisfies the equation

$$\slash v h_v=h_v .\eqno(41)$$

These results
follow from a strong interaction effective Lagrangian
which is the starting point of HQET:

$${\cal L} = \sum_v {\cal L}_{h_v} =   \sum_v i \spc
\bar h_v v^\mu D_\mu h_v \hskip 1cm .\eqno(42)$$

The effective Lagrangian (42) has two symmetries that are not
present in the original QCD lagrangian  and are a consequence of the heavy
quark
mass limit:
\item {1)} {$SU(2)$ spin symmetry, which follows from the absence of $\gamma$
matrices in (42).}
\item { 2)} {$SU(N)$ ($N=2$ if only charm and beauty are considered) flavour
symmetry, which follows from the independence of
the lagrangian from the heavy quark flavour.}

Similar to Eq.(2) relating $f_B$ to $f_D$,
because of these symmetries one
can relate several physical quantities that are in principle independent.
In particular, all
the form factors in Eqs.(34-36) can be expressed in terms of
one universal form factor $\xi$ (Isgur-Wise function). $\xi$ is a function
of $w=v\cdot v'$, where $v$ and $v'$ are the $B$ and $D$ four velocities:

$$\xi=\xi(w)\eqno(43)$$
with

$$w=v \cdot v' ={m^2_B+m^2_D-q^2 \over 2m_B m_D} \spc .\eqno(44)$$
Notice that $\xi(1)=1$ since the vector charge is the generator of
the flavour symmetry.

The matrix elements in (34-36) are expressed in terms of $\xi$ (neglecting
$1/m_Q$
corrections) as follows:

$$<D(v')|V^\mu|B(v)>={\sqrt{m_B m_D}} \spc \xi(w) \spc (v+v')^\mu\eqno(45)$$

$$\eqalign{
<D^*(v', \epsilon)|V^\mu-A^\mu|B(v)> &=\spc
{\sqrt{m_B m_D}}\spc \xi(w)\Big [ i \epsilon^{\mu \nu \alpha \beta}
\epsilon^*_{\nu} v_{\alpha} v'_{\beta}\cr
& - (1+w)\epsilon^{\mu *} +
(\epsilon^* \cdot v) v'^\mu \Big ]\spc .\cr
}\eqno(46)$$
\par
The relations among the form factors in Eqs.(34-36)
and the universal function
$\xi$ can be obtained by a straightforward calculation.
\par
We shall also consider the semileptonic transitions
$$B \to D^{**} \ell \nu \spc ,\eqno(47)$$
\par\noindent
where $D^{**}$ are positive parity $p$-wave charmed mesons.
According to the quark model one expects four states:
$^{2S+1}L_{J}= \spc {^3P_2}, \spc {^3P_1}, \spc {^3P_0}$
(corresponding to total spin $S=1$)
and  $^1P_1$ (corresponding to total spin $S=0$). In the HQET
the heavy quark spin $s_Q$ is decoupled from
the light quark angular momentum
$\vec s_\ell= \vec J - \vec s_Q$, so that
it is convenient to work with the two degenerate multiplets
$J^P=(0^+_{1/2},1^+_{1/2})$ and
$J^P=(1^+_{3/2},2^+_{3/2})$ which differ by the value
$s_\ell=1/2$ and $3/2$ respectively (the relative orbital momentum is $L=1$).
The $1^+$ states
defined above are given by the following linear combinations
[26,28]:
$$\eqalign{
|1^+_{3/2}>&=\sqrt{{2\over 3}}\ |^1P_1>+\sqrt{{1\over 3}}\ |^3P_1>\cr
|1^+_{1/2}>&=\sqrt{{1\over 3}}\ |^1P_1>-\sqrt{{2\over 3}}\ |^3P_1>\cr
}\eqno(48)$$
while
$|0^+_{1/2}>=|^3P_0>$ and $|2^+_{3/2}>=|^3P_2>$.
The $2^+_{3/2}$ state has been observed experimentally  with a mass
of $2460 \hskip 3pt MeV$: it is denoted by $D_2^*(2460)$.
The $1^+_{3/2}$ meson can be
identified with $D_1(2420)$ even though a
$1^+_{1/2}$ component
 is likely to be contained in this state [26,29].
They are both narrow
$(\Gamma \leq 20 \hskip 3pt MeV)$
since their strong decays occur in $d$-wave, in contrast with the states
$D_0$ (the $0^+_{1/2}$ state) and $D'_1$ (the $1^+_{1/2}$ state)
that can also decay in
$s$-wave.
\par
In the HQET the matrix elements for the decays (47) can be written
in terms of two universal functions $\tau_{1/2}(w)$ and $\tau_{3/2}(w)$ [27].
The form factor
$\tau_{3/2}(w)$ describes the transitions into the $2^+_{3/2}$ and $1^+_{3/2}$
states, and $\tau_{1/2}(w)$ those to the $1^+_{1/2}$
and $0^+_{1/2}$ states.
The hadronic matrix elements for the decays (47) in the $m_Q \to \infty$ limit
can be written as [27]:

$$ <D_0(v')|A_\mu|B(v)> = - 2 {\sqrt {m_B m_{D^{**}}}}  \hskip 4pt
\tau_{1/2} (w)  \hskip 4pt (v-v')_\mu \eqno(49)$$

$$\eqalign{
<D'_1(v',\epsilon)|&V_\mu - A_\mu|B(v)> = {\sqrt {m_B m_{D^{**}}}}
 \hskip 4pt \tau_{1/2}(w) \times \cr
&\times [2 \hskip 4pt (w -1) \hskip 4pt \epsilon^*_\mu
+ 2 \hskip 4pt (\epsilon^*\cdot v) v'_\mu -
i \epsilon_{\mu\nu\alpha\beta} \hskip 4pt \epsilon^{*\nu}
(v+v')^\alpha (v-v')^\beta] }\eqno(50)$$

$$\eqalign{
<D_1(v',\epsilon)|&V_\mu - A_\mu|B(v)> = {\sqrt {m_B m_{D^{**}}}}
 \hskip 4pt \tau_{3/2}(w) \times \cr
&\times \Big \{ [{ (1 - w^2 ) \over \sqrt 2} \hskip 4pt
\epsilon^*_\mu
+ { (\epsilon^*\cdot v) \over \sqrt 2} [- 3 \hskip 4pt v_\mu +
 (w -2) \hskip 4pt v'_\mu ]  \cr
&+
i \hskip 4pt {(w - 1) \over 2 \sqrt 2} \epsilon_{\mu\nu\alpha\beta}
\epsilon^{*\nu} (v+v')^\alpha (v-v')^\beta] \Big \} }\eqno(51)$$

$$\eqalign{
<D^*_2(v',\epsilon)|&V_\mu - A_\mu|B(v)> = {\sqrt {m_B m_{D^{**}}}}
 \hskip 3pt \tau_{3/2}(w) \hskip 3pt \cr
&[ \hskip 3pt i \hskip 4pt {{\sqrt 3} \over 2} \epsilon_{\mu\alpha\beta\gamma}
\epsilon^{*\alpha\nu} v_\nu (v+v')^\beta (v-v')^\gamma
- {\sqrt 3} (w +1) \epsilon^*_{\mu\alpha} v^\alpha \hskip 3pt +\cr
& {\sqrt 3} \epsilon^*_{\alpha\beta} v^\alpha v^\beta v'_\mu ]
}\eqno(52)$$
\par\noindent
where now $V^\mu= {\bar c}_{v'} \gamma^\mu b_v$,
$A^\mu= {\bar c}_{v'} \gamma^\mu \gamma_5 b_v$,
$c_{v'}$ and  $b_v$ are charm and beauty quark operators in HQET,
and $m_{D^{**}}$
is the mass of the
positive parity charmed resonances.
In the following Section we shall review the QCD sum rules calculation
of $\xi(w)$, $\tau_{1/2}(w)$ and $\tau_{3/2}(w)$.
\bigskip
\noindent {\bf 5. Three point functions}
\medskip
To obtain QCD sum rules for the form factors one starts from the three-point
correlators [30]:
$$\Pi_{\Gamma,\mu}(p,p^{\prime},q)=i^2\int dx \hskip 2pt dy \hskip 2pt
e^{i(p'\cdot x-p\cdot y)}
<0|T \left(J_{\Gamma}(x) J^{V,A}_{\mu}(0) J_5(y)\right)|0>\hskip 5pt.
\eqno(53)$$
In (53) $J_{\Gamma}$ denotes the currents interpolating the final charmed
meson with $\Gamma$ the appropriate set of Lorentz indices
(e.g. $J_{\Gamma}={\bar q}c$,
${\bar q}i\gamma_5c$ and
$\bar q(\gamma_{\nu} {\overleftarrow {\overrightarrow \partial}}_\mu+
\gamma_{\mu}
 {\overleftarrow {\overrightarrow \partial}}_\nu -
{1 \over 2} g_{\mu\nu} \gamma^\rho
 {\overleftarrow {\overrightarrow \partial}}_\rho)c$
for $0^{\pm}$ and $2^+$
respectively), and $J_5$ is the
analogous operator for the initial $B$-meson, see Eq.(3).
The form factors are extracted by expanding, in each case, Eq.(53) in
independent Lorentz structures.
\par
Similar to the analysis of two-point functions, we write
a sum rule by computing (53) in two different ways: by including hadronic
states plus a continuum of states modelled by perturbative QCD on one side,
and by
performing an OPE in the framework of QCD on the other side,
accounting for the asymptotic freedom contribution plus higher dimensional
non-perturbative condensates. Finally,
these alternative representations
are matched to one another, and the convergence is improved by one of the
methods discussed in Sect.2, {\it e.g.}
by taking the Borel
transform of both sides of the sum rule. The double Borel
transform, as applied to any of the Lorentz invariant functions
$\Pi_i(p^2,p'^2,q^2)$ appearing on the left hand side of (53), looks as
follows:
\vskip 5pt
$$\Pi_i(\sigma, \sigma',q^2)= - {1 \over 4 \pi^2} \int  \rho_i(s, s', q^2)
\hskip 3pt \exp(-s \sigma ) \hskip 3pt \exp(-s' \sigma')
\spc ds \hskip 3pt ds' \spc \spc ,
\eqno(54)$$
\vskip 10pt
\par\noindent
where
$\Pi_i(\sigma, \sigma',q^2)$
is obtained applying the Borel transform to the
OPE of
$\Pi_i(p^2,p'^2,q^2)$,
and the double spectral function $ \rho_i(s, s', q^2)$ is saturated by the
lowest hadronic states, involving the form factors of interest, plus the QCD
continuum starting at some thresholds $s_0$, $s'_0$.
\par
In order to perform the heavy-quark mass limit and then estimate the
dependence of the the form factors on $w$,
we estimate the sum
rules at $q^2=0$, and take the limit $m_Q\equiv m_b\to\infty$ with
$m_c=m_b/\sqrt Z$, $Z$ fixed [31].
The form factors then become functions of $w$
{\it via} Eq.(44) at $q^2=0$, which reads:
$$\sqrt Z=w+\sqrt{w^2-1}\hskip 4pt.\eqno(55)$$
\par\noindent
In the
case of $0^- \to (0^-, 1^-)$ semileptonic transitions, this procedure
reproduces the main features of the results obtained by
directly working in the infinite heavy-quark limit [17,32].
As for the Borel parameters
$\sigma$ and $\sigma'$, in the heavy-quark limit, in
analogy with (12), we write
$\tau_1=  \hskip 2pt 2 \hskip 2pt m_b \hskip 2pt \sigma$ and
$\tau_2=  \hskip 2pt 2 \hskip 2pt m_c \hskip 2pt \sigma'$.
For the continuum thresholds we put
$s_0'=m_b^2/Z+2 \hskip 2pt  m_b \hskip 2pt \mu_+/\sqrt Z$,
$s_0=m_b^2+2 \hskip 2pt m_b \hskip 2pt \mu_-$ and
for the meson masses we introduce the "binding energies" as follows:
$m_B^2=m_b^2+2 \hskip 2pt m_b \hskip 2pt {\omega_R}$,
$ m_P^2=m_b^2/Z+2 \hskip 2pt m_b{\omega_R}^{\prime}/\sqrt Z$
where $P=D,D^*,D^{**}$.
\par
The resulting sum rules for the universal form factors are [33]:
$$\eqalignno{
\xi(w)&={e^{(\omega_R \tau)}\over{\displaystyle {\hat F}^2 }}
\Big\{ {3\over 4\pi^2}{Z\over(\sqrt Z+1)^3}H(w)\cr
&-<\bar q q>\left[1-{\lambda^2 \tau^2 (w-1)\over 48}\right]
e^{-\lambda^2 \tau^2 (w+1)/16)} \Big\}&(56)\cr}$$
$$\eqalignno{
\tau_{1/2}(w)&=
{e^{(\omega_R \tau_1 + \omega_R^\prime \tau_2)}\over
f^{(+)}{\hat F} }\Big\{{3\over 8\pi^2}
{Z\over {\sqrt Z}+1}J(w)\cr
&-{<\bar q q> \over 2} \left[1-{\lambda^2 \tau_1 \tau_2\over 12}
(w+1)\right]
\hskip 3pt
e^{-(\lambda^2 (\tau^2_1/8 +\tau_2^2/8 + 2w\tau_1 \tau_2 /8))}\Big\}&(57)\cr}$$
$$\tau_{3/2}(w)=
{e^{(\omega_R \tau_1 + \omega_R^\prime \tau_2)}\over f_T
{\hat F}} {\sqrt 3\over \pi^2}
{Z^{3/2}\over (\sqrt Z +1)^5} I(w) \hskip 5pt,\eqno(58)$$

\par\noindent
where $\lambda^2=M_0^2/2$ arises from the contribution of
the dimension 5 operator introduced with regard to Eqs.(9) and (11).
In Eq.(56) we have set $\tau_1=\tau_2=\tau/2$ and $\omega_R=\omega'_R$,
which ensures $\xi(1)=1$ for any value of $\tau$.
Actually, once $\hat F$ is replaced by the sum rule (19), the dependence on
$\omega_R$ drops from the ratio, and this minimizes the sensitivity of $\xi(w)$
on the values of this quantity.
Explicit expressions for $O(\alpha_s)$ corrections to (56) have been presented
in [34]. The complete $\alpha_s$ corrections for the case of (57) and
(58) are not available yet.
\par
In (57), (58) we use the leptonic constants $f^{(+)}$ and $f_T$ defined as
follows:

$${\displaystyle{<0|V_{\mu}|M(0^+_{1/2};p)>=i{f^{(+)}\over\sqrt{m_Q}}p_{\mu}}}
\hskip 5pt,\eqno(59)$$
$${\displaystyle{<0|J_{\mu\nu}|M(2^+_{3/2};p,\epsilon)>=
\epsilon_{\mu\nu}f_T\sqrt{m_Q}}}\hskip 5pt.\eqno(60)$$

The numerical values for $f_T$ and $f^{(+)}$ are obtained,
in analogy with ${\hat F}$, by two-point  QCD sum rules [33] and the results
are as follows:
$f_T=(0.18 \pm 0.02) \spc GeV^{5\over 2}$ and $f^{(+)}=(0.46 \pm 0.06)\spc
GeV^{3 \over 2}$.

The functions $H(w)$, $J(w)$
and $I(w)$ in (56)-(58) are given by:
$$H(w)={1\over\sqrt Z-1}\int_0^{2\mu_-}dy\int_0^{2\mu_-}dy^{\prime} \hskip 3pt
(y+y^{\prime}) \hskip 3pt e^{-\tau (y+y^{\prime})/4}
\hskip 2pt \Theta (2 \hskip 2pt w \hskip 2pt y \hskip 2pt y^{\prime}-y^2
-y^{\prime 2})\eqno(61)$$

$$\eqalignno{J(w)={1\over (\sqrt Z-1)^3}\int_0^{2\mu_+}dy\int_0^{2\mu_-}dy^
{\prime} \hskip 3pt&
(y-y^{\prime}) e^{-(y \tau_2 /2+y^\prime \tau_1/2)}\times\cr
&\Theta (2 \hskip 2pt w \hskip 2pt y \hskip 2pt y^{\prime}
-y^2-y^{\prime 2})&(62)\cr}$$

$$\eqalignno{
I(w)={1\over (\sqrt Z-1)^3}\int_0^{2\mu_+}&dy\int_0^{2\mu_-}dy^{\prime}
[(y^2+2yy^\prime)(Z-\sqrt Z+1)-3\sqrt Z y^{\prime 2}] \times\cr
&e^{-(y \tau_2 /2+y^\prime \tau_1/2)}\hskip 2pt
\Theta (2 \hskip 2pt w \hskip 2pt y \hskip 2pt y^{\prime}-y^2
-y^{\prime 2})\hskip 5pt,&(63)\cr}$$
\par\noindent
where $\mu_\pm$ are the effective thresholds in the heavy quark limit.

The stability analysis of the results for the form factors is
an obvious extension of the method described in Sect.2. One observes
explicitly
that  the non-perturbative terms turn out to be
numerically negligible for $\tau_{3/2}$, and therefore
have not been included in (58). Moreover, it should be
 noted
that
the optimal values of the Borel parameters $\tau_1$ and $\tau_2$ as well as the
continuum thresholds $\mu_+$ and $\mu_-$ are taken at
the same values obtained
in the analysis of two-point functions [33], in particular
$\mu_- = 0.7 \pm 0.1 \spc GeV$ and
$\mu_+ = 1.3 \pm 0.1 \spc GeV$. Finally,
the form factors $\tau_{1/2}$ and $\tau_{3/2}$ should include a multiplicative
$O(\alpha_s)$ correction which depends on the anomalous dimension of the
current operators and on $w$. Its numerical effect is
small and can be included in the theoretical uncertainties of the results.
Furthermore,
as far as $\xi$ is concerned, we notice that the large $O(\alpha_s)$
corrections affecting ${\hat F}$ (see discussion in Section 2) should
drop from the ratio in (56), which also preserves $\xi(1)=1$.
\par
An important role in the numerical analysis of $\xi$ is played by the
continuum.
In spite of the denominators $(\sqrt Z-1)^n=
(w-1+\sqrt{w^2-1})^n$, with $n=1,3$, the integrals $H(w)$, $J(w)$ and
$I(w)$ are finite for $w\to 1$.
On the other hand the slope of the universal function $\xi$
turns out to be divergent, a result
which is unnatural. As shown in [17] this is an artifact of the
approximations used in modelling the higher mass resonances as a continuum of
states. To cure this problem for the ``Isgur-Wise'' function $\xi(w)$ one
can modify the integration region in (61) without affecting the
normalization constraint at zero recoil ${\xi(1)=1}$
(which is easily seen to be verified by
the sum rule). One can replace, for
example, the $\Theta$ function by
$\Theta (2\hskip 2pt w \hskip 2pt y \hskip 2pt y^\prime-y^2-y^{\prime 2})
\times\Theta[2\mu(1+w-\sqrt{w^2-1}-y-y^\prime]$ (case a) or,
as another possibility,
one can just change the upper limit of integration
$2\mu\to 2\mu(1+w+\sqrt{w^2-1})/(1+w)$ (case b). Both procedures
lead to a finite
slope. However, they add some significant uncertainty to the determination of
$\xi(w)$ for values of $w$ relevant to nonzero recoil, and
in particular by varying the continuum integration domain
in
Eq.(56) one obtains different values for the slope
$\xi'(1)$. The value that better fits the experimental data
corresponds to the case a and is [17]:

$$\xi'(1)= -1.28 \pm 0.25 \spc ,\eqno(64)$$
\par\noindent
while in [35] it is suggested to adopt the case b, which leads to:

$$\xi'(1)= -0.70 \pm 0.10 \spc .\eqno(65)$$
\par\noindent
An analogous result has been found
in [34] including $O(\alpha_s)$ corrections.
We observe that all these values satisfy the lower bound:

$$|\xi'(1)| \ge 0.25 \eqno(66)$$
\par\noindent
obtained in [36].
\par
Regarding the
numerical evaluation of $\tau_{1/2}(w)$ and $\tau_{3/2}(w)$ according to
Eqs.(57)-(58), we notice that, in this case,
there is no general normalization condition on the form factors
at $w=1$. In Fig.1
we depict the functions $\tau_{1/2}(w)$ and $\tau_{3/2}(w)$ for the central
values of $\mu_+$ and $\mu_-$ and with $1/\tau_1\simeq 1/\tau_2\simeq
1.5\ GeV$.
We observe that the numerical values at $w=1$:
$\tau_{1/2}(1) \simeq \tau_{3/2}(1) \simeq 0.22$ are in reasonable agreement
with the outcome of the constituent quark model in [27].
Moreover, they also agree with the estimate
$\tau_{1/2}(1) \simeq 0.35 \pm 0.2$
given in [35]. Finally,  we observe that the values in Fig.1 are inside
the upper bound put by the optical sum rule [37]:

$$|\tau_{1/2}(1)|^2 + 2 |\tau_{3/2}(1)|^2 < {1 \over 2}{m_D-m_c\over
m_{D^{**}}-m_D}\simeq 0.5 \spc .\eqno(67)$$

In Table 2 we list the values of the branching ratios resulting
from the analysis presented in this Section.
\vskip 1cm
\baselineskip 14pt
\setbox5=\vbox{\hsize 23.pc \noindent \strut
{$$\vbox{\settabs
\+ Channel \hskip 60pt & Branching Ratio \hskip 30pt \cr
\+ Channel  & Branching Ratio \hskip 30pt \cr
\+ $B \to D(0^-) \hskip 2pt \ell \hskip 2pt \nu$ &
$1.4 \times 10^{-2} \hskip 4pt \Big({V_{cb} \over 0.04}\Big)^2$ \cr
\+ $B \to D(1^-) \hskip 2pt \ell \hskip 2pt \nu$ &
$4.3 \times 10^{-2} \hskip 4pt \Big({V_{cb} \over 0.04}\Big)^2$ \cr
\+ $B \to D_0 \hskip 2pt \ell \hskip 2pt\nu$ &
$5 \times 10^{-4} \hskip 4pt \Big({V_{cb} \over 0.04}\Big)^2$ \cr
\+ $B \to D'_1 \hskip 2pt \ell \hskip 2pt \nu$ &
$7 \times 10^{-4} \hskip 4pt \Big({V_{cb} \over 0.04}\Big)^2$ \cr
\+ $B \to D_1 \hskip 2pt\ell \hskip 2pt \nu$ &
$1 \times 10^{-3} \hskip 4pt \Big({V_{cb} \over 0.04}\Big)^2$ \cr
\+ $B \to D_2^{**} \hskip 2pt \ell \hskip 2pt \nu$ &
$2 \times 10^{-3} \hskip 4pt \Big({V_{cb} \over 0.04}\Big)^2$ \cr
}$$ } }
$$\boxit{\box5}$$
\hsize=146truemm
\baselineskip 14pt
\hoffset=0.141in
\par\noindent {\small {\bf Table 2 }. Branching ratios obtained for the
various B-meson semileptonic
transitions in the heavy-quark limit [33]
($\tau_B=1.21 \hskip 3pt ps$).}
\bigskip
\noindent{\bf 6. Semileptonic form factors for finite heavy quark masses}
\medskip
The method of QCD Sum Rules can be applied to the evaluation of the form
factors in Eqs.(34-36) without performing the limit $m_Q \to \infty$.
This offers the possibility to evaluate systematically $O(1/m_Q)$
corrections for the transitions $B \to D, D^*$.
We shall consider also the form factors for decays into the
positive parity $0^+_{1/2}$ and $1^+_{1/2}$ states
(the other form factors have not been computed yet).
The form factors are given by:

$${1 \over i} \hskip 3pt <D_0^{**}(p')|\spc {\bar c}\gamma_\mu
\gamma_5b|\spc B(p)> =
\spc g_+ \spc (p'+p)_\mu + g_- \hskip 3pt  q_\mu\eqno(68)$$
$${1 \over i} \hskip 3pt <D^\prime_1(p',\epsilon)|\spc {\bar c}\gamma_\mu
\gamma_5b|\spc B(p)> =
\spc i \spc f \spc \epsilon_{\mu\nu\alpha\beta} \epsilon ^{*\nu}
(p'+p)^\alpha q^\beta \eqno(69)$$
$${1 \over i} \hskip 3pt <D^\prime_1(p',\epsilon)|\spc {\bar c}\gamma_\mu
b|\spc B(p)> =
\spc f_0 \spc \epsilon^*_\mu + f_+ \spc (\epsilon^* \cdot p) (p'+p)_\mu +
f_- \spc (\epsilon^* \cdot p) q_\mu \eqno(70)$$
\par\noindent
where $\epsilon ^\mu$ is the $D^\prime_1$ polarization vector and the
form factors $g_\pm$, $f$, $f_0$, $f_\pm$ are functions of $q^2$. For
this application it is also necessary
to evaluate  {\it via} QCD sum rules the vacuum-to-meson constants:
$$ <0| {\bar q} c| D_0> = {f_{D_0} \hskip 3pt
m^2_{D_0} \over m_c}
\eqno(71)$$
\par\noindent and
$$ <0| {\bar q} \gamma_\mu\gamma_5 c| D^\prime_1(p,\epsilon)> =
{m^2_{D^\prime_1} \over g_{D^\prime_1}} \epsilon_\mu(p) \hskip 5pt .
\eqno(72)$$
\par\noindent
The result for the mass is $m_{D_0}= m_{D^\prime_1}=2.5\pm 0.1 \spc GeV$,
and the leptonic constants are given in Table 3 [38].
\setbox6=\vbox{\hsize 22.pc \noindent \strut
{$$\vbox{\settabs
\+ Leptonic constant \hskip 60pt & Theoretical result \cr
\+ Leptonic constant \hskip 60pt & Theoretical result \cr
\+ $\hskip 20pt f_{D_0} \hskip 30pt $& $\hskip 10pt 170\pm20
\hskip 10pt MeV$\cr
\+ $\hskip 20pt g_{D'_1} \hskip 30pt $& $\hskip 10pt 9.8\pm1.5 $\cr
}$$ } }
$$\boxit{\box6}$$
\centerline{\small {\bf Table 3} Leptonic constants, see Eqs. (71), (72).}
\vskip 1cm
\hsize=146truemm
\baselineskip 14pt
\hoffset=0.141in
\par
Our results for the values of the various form factors at
$Q^2=0$  are reported
in Table 4. As for the $Q^2$-dependence of the form factors,
it could be derived by QCD sum rules; however,
for simplicity,
we assume a simple pole behaviour
$\displaystyle F_i(Q^2)=F_i(0)/(1+ {Q^2\over m^2})$, which is adequate
in the present case. This is also
suggested by dispersion relations,
where the pole masses
are those of $b { \bar c}$ mesons with
appropriate quantum numbers.
\par
%
%\hoffset=-2.2 truecm
%
%
\setbox7=\vbox{\hsize 27.pc \noindent \strut
{$$\vbox{\settabs
\+ $B \to D(0^-)$ \hskip 20pt & $g_+(0)$ \hskip 40pt &
$f(0)$ \hskip 40pt & $f_+(0)$ \hskip 40pt &
$f_0(0)$ \cr
\+ & $g_+(0)$\hskip 10pt & $f(0)$ \hskip 10pt & $f_+(0)$  \hskip 10pt &
$f_0(0)$  \hskip 10pt \cr
\+ & $G_+(0)$\hskip 10pt &
$F(0)$\hskip 10pt & $F_+(0)$\hskip 10pt &
$F_0(0)$\hskip 10pt \cr
\+ & & ($GeV^{-1}$)& ($GeV^{-1}$)&($GeV)$\cr
\+ & && &\cr
\+ $B \to D(0^-)$ \hskip 20pt & $0.67\pm0.19$&&&\cr
\+ $B \to D^*(1^-)$ \hskip 20pt & &$0.12\pm0.02$&
$-0.05\pm0.02$&$4.35\pm0.15$\cr
\+ $B \to D_0^{**}(0^+)$ \hskip 20pt & $0.30\pm0.03$&&&\cr
\+ $B \to D^{**}(1^+)$ \hskip 20pt & &$0.10\pm0.03$&
$-0.02\pm0.02$&$0.9\pm0.3$\cr
}$$ } }
$$\boxit{\box7}$$
%\vskip 5cm
\hsize=146truemm
\baselineskip 14pt
\hoffset=0.141in
\centerline {\small {\bf Table 4} Form factors for B decays for finite $m_Q$}
%\vskip 1cm
\vfill\eject
\par
The results for
the widths are rather insensitive to the precise value of these masses;
we
take $m = 6.34 \hskip 3pt GeV$ for the  $1^-$ pole (for
 $G_+, F, f_0$ and $f_+$) and $m = 6.73 \hskip 3pt GeV$
for the $1^+$ pole (for $g_+, f, F_0$ and $F_+$).
\par
{}From the values in Table 4 we can compute the branching ratios
for the decays $B \to D, D^*$, $B \to D_0, D'_1$. The results are collected
in  Table 5.
\vskip 1cm
\setbox8=\vbox{\hsize 26.pc \noindent \strut
{$$\vbox{\settabs
\+ Decay mode \hskip 10pt & BR (Theory )\hskip 40pt &
BR (Exp. [39]) \hskip 30pt \cr
\+ Decay mode & \hskip 10pt BR (Theory )\hskip 20pt &
\hskip 20 pt BR (Exp. [39])  \cr
\+ & &\cr
\+ $B \to D \ell \nu $ \hskip 20pt & $1.5 \hskip 2pt ({V_{cb} \over 0.04})^2
\times 10^{-2}$&
$1.75\pm 0.42\pm0.35 \times 10^{-2}$\cr
\+ $B \to D^* \ell \nu$ \hskip 20pt & $4.6 \hskip 2pt ({V_{cb} \over 0.04})^2
\times 10^{-2}$&
$4.8\pm 0.4\pm0.7 \times 10^{-2}$\cr
\+ $B \to D_0 \ell \nu$ \hskip 20pt & $0.15 \hskip 2pt ({V_{cb} \over 0.04})^2
\times 10^{-2}$&\cr
\+ $B \to D'_1 \ell \nu$ \hskip 20pt & $0.15 \hskip 2pt ({V_{cb} \over 0.04})^2
\times 10^{-2}$&\cr
}$$ } }
$$\boxit{\box8}$$
\par
\centerline {\small {\bf Table 5} Branching ratios for finite $m_Q$}
%\vskip 1cm
%
\hoffset=0.141in
\voffset=0.33in
\magnification=\magstep1
\hsize=146truemm
\baselineskip=14pt
\vskip 1cm
\par
Comparing these results with those in Table 2, we observe that the $1/m_Q$
corrections affect much more strongly the decays into positive parity charmed
states.
This can be qualitatively understood by recalling that in the case of higher
charmed resonances the expansion parameter can be expected to be
$\displaystyle \propto (m_{D^{**}} - m_c)/m_c$, which is a number of order one.
For the transitions into
$D,D^*$ the corrections linear in $1/m_Q$
are forbidden in the most relevant form factors at $w=1$ [40] and this seems to
protect the leading order result.
Actually, the systematic analysis of $B$ decays into negative parity states at
the next-to-leading order in $1/m_Q$ implies the introduction of four new
functions, in addition to $\xi$ [40]. Some of these functions are also subject
to normalization conditions at the zero recoil point $w=1$. They have been
recently analyzed in the framework of QCD sum rules in [41].
\bigskip
\noindent {\bf 7. Concluding remarks}
\medskip
\par
Summarizing the results presented in the previous sections, QCD sum rules
provide a conceptually consistent (and numerically simple) method to estimate
from first principles semileptonic and purely leptonic transitions of the $B$
meson.
\par
For the case of $f_B$, predicted values are in the potential of a $B$ factory,
and indeed such a measurement would be really welcome for theoretical reasons.
\par
For $B \to D,D^*$ semileptonic transitions, the method can provide a
systematic analysis of the relevant form factors, both for finite and for
infinite $m_b$. In the last regard, the results (and the simplicity) of the
HQET can be easily incorporated in this framework, and the Isgur-Wise function
$\xi(w)$ can be predicted, naturally accomodating the zero-recoil normalization
$\xi(1)=1$. Order $\alpha_s$ effects as well as the form factors relevant to
the general next-to-leading $1/m_Q$ structure can be also accounted for. Thus
the situation with the transitions to negative-parity charmed states looks
well settled from the theoretical point of view, waiting for more
accurate experimental  data,
which would allow a precise and model independent
determination of $V_{cb}$.
\par
Also the $B \to D^{**}$ semileptonic transitions can be evaluated in the
framework of QCD sum rules, where $D^{**}$ represents the orbitally excited
charmed resonances. Indeed, the analysis of these decays is theoretically
related to the previous ones in this framework, which therefore provides a
unified treatment of all form factors in terms of the same fundamental
parameters. Moreover, the case of $D^{**}$ final states can be discussed in the
view of the heavy quark mass limit and the two universal functions
$\tau_{1/2}(w)$  and
$\tau_{3/2}(w)$ can be predicted. The current analysis, more difficult as it
involves operators with derivatives of the quark fields, is not complete as it
is in the case of $\xi(w)$, in particular the order $\alpha_s$ effects are not
taken into account yet. In addition, the $\tau$'s are not normalized at zero
recoil so that, in general, uncertainties are larger and predictions can be
considered as more qualitative. Nevertheless, it is encouraging that the values
of $\tau_i(1)$ thus found are in reasonable agreement with the independent
quark model estimates, and comply with the theoretical upper bound from the
optical sum rule.
\par
Concerning the branching ratio of $B \to D^{**}$, the results indicate
values of the order of
$ a \spc few \times 10^{-3}$ for the sum of the various p-wave resonances,
which is not enough to fill the observed gap between
the inclusive semileptonic rate and the sum
of the exclusive rates $B \to D,D^* \ell \nu$.
Clearly, this issue deserves further theoretical attention and represents an
interesting point for the physics at $B$ factories.
\vfill\eject
%\bigskip
\noindent{\bf References}
\medskip
\item{[1]} M.A.Shifman, A.I.Vainshtein and V.I.Zakharov, Nucl. Phys.
{\bf B147} (1979) 385, 448.
\medskip
\item{[2]} L.J.Reinders, H.R.Rubinstein and S.Yazaki, Nucl.Phys. {\bf
B186} (1981) 109.
\medskip
\item{[3]} ARGUS collaboration, H.Albrecht et al., Phys. Lett. {\bf B197}
 (1990) 452; {\bf B229} (1989) 175.
\medskip
\item{[4]} CLEO collaboration,D.Bortoletto et al., Phys. Rev. Lett. {\bf 63}
(1989) 1667.
\medskip
\item{[5]} For a review see H.Georgi, report HUTP-91-A039.
See also T.Mannel, this report.
\medskip
\item{[6]} M.Lusignoli, L.Maiani, G.Mar\-ti\-nel\-li and L.Re\-i\-na,
Nucl. Phys. {\bf 369} (1992) 139.
\medskip
\item{[7]} Particle Data Group, Review of Particle Properties, Phys. Rev.
{\bf D45} (1992) S1.
J.Adler {\it et al.}, Phys. Rev. Lett. {\bf 60} (1988) 1375.
\medskip
\item{[8]} S.Aoki {\it et al.}, CERN-PPE/92-157 (1992). This result is
compatible with the determinations obtained by the non leptonic $B$ decays
using factorization, see {\it e.g.}: D.Bortoletto and S.Stone, Phys. Rev.
Lett. {\bf 65} (1990) 1951 and H.Albrecht et al., ARGUS Collab., Zeit. Phys.
{\bf C54} (1992) 1.
\medskip
\item{[9]} D.J.Broadhurst and S.C.Generalis, Open University Report No.
OUT-4102-8/R (1982);
S.C.Generalis, Ph.D.Thesis, Open University Report No. OUT-4102-13 (1984).
See also K.Schilcher, M.D.Tran and N.F.Nasrallah, Nucl. Phys. {\bf B181}
(1981) 91.
\medskip
\item{[10]} T.M.Aliev and V.L.Eletsky, Sov. J. Nucl. Phys. {\bf 38} (1983)
936.
\medskip
\item{[11]} K.Schilcher and Y.L.Wu, Zeit. Phys. {\bf C54} (1992) 163.
\medskip
\item{[12]} C.A.Dominguez and N.Paver, Phys Lett. {\bf B197} (1987) 423;
(E) {\bf B199} (1987) 596.
\medskip
\item{[13]} S.Narison, Phys. Lett. {\bf B198} (1987) 104.
\medskip
\item{[14]} L.J.Reinders, Phys. Rev. {\bf D38} (1988) 423.
\medskip
\item{[15]} C.A.Dominguez and N.Paver, Phys. Lett. {\bf B269} (1991) 169.
\medskip
\item{[16]} E.Shuryak, Nucl. Phys.  {\bf B198} (1982) 83;
V.Eletsky and E.Shuryak, Phys. Lett. {\bf B276} (1992) 191.
\medskip
\item{[17]} M.Neubert, Phys. Rev. D {\bf 245} (1992) 2451.
\medskip
\item{[18]} R.A.Bertlmann, Nucl. Phys. {\bf B204} (1982) 387.
\medskip
\item{[19]} E.Bagan, P.Ball, V.M.Braun and H.G.Dosch, Phys. Lett.
{\bf B278} (1992) 457.
\medskip
\item{[20]} G.P.Korchemsky and A.V.Radyushkin,
Phys. Lett. {\bf B279} (1992) 359.
\medskip
\item{[21]} D.J.Broadhurst and A.G.Grozin, Phys. Lett. {\bf B267}
(1991) 105; {\bf B274} (1992) 421.
\medskip
\item{[22]} C.R.Allton, C.T.Sachrajda, V.Lubicz, L.Maiani and
G.Mar\-ti\-nel\-li, Nu\-cl. Phys. {\bf B349} (1991) 598. For a review see
G.Martinelli, this report.
\medskip
\item{[23]} M.B.Voloshin, Sov. J. Nucl. Phys. {\bf 29} (1979) 703.
\medskip
\item{[24]} C.A.Dominguez and N.Paver, Phys. Lett. {\bf B293} (1992) 197.
\medskip
\item{[25]} J. Schwinger, {\it Particles, Sources and Fields}, Vol.II,
(Addison-Wesley, 1973).
\medskip
\item{[26]} J.L.Rosner, Comm. Nucl. Part. Phys. {\bf 16} (1986) 109.
\medskip
\item{[27]} N.Isgur and M.B.Wise, Phys. Rev. {\bf D43} (1991) 819.
\medskip
\item{[28]} S.Balk, J.G.K\"orner, G.Thomson  and F.Hussain,
report IC/92/397.
\medskip
\item{[29]} U.Kilian, J.G.K\"orner and D.Pirjol,
Phys. Lett. {\bf B288} (1992) 360; A.F.Falk and M.Luke, Phys. Lett. {\bf B292}
(1992) 119.
\medskip
\item{[30]} B.L.Ioffe and A.V.Smilga, Phys. Lett. {\bf B114} (1982) 353;
Nucl. Phys. {\bf B216} (1983) 373.
\item{[31]} M.Neubert and V.Rieckert,
Nucl. Phys. {\bf B382} (1992) 97.
\medskip
\item{[32]} A.V.Radyushkin,
Phys. Lett. {\bf B271} (1991) 218.
\medskip
\item{[33]} P.Colangelo, G.Nardulli and N.Paver,
Phys. Lett. {\bf B293} (1992) 207.
\medskip
\item{[34]} E.Bagan, P.Ball and P.Gosdzinsky,
report UAB-FT-292 (1992).
\medskip
\item{[35]} B.Blok and M.Shifman, Report TPI-MINN-92-93/T (1992).
\medskip
\item{[36]} J.D.Bjorken, report SLAC-PUB-5278 (1990).
\medskip
\item{[37]} M.Voloshin, Phys. Rev {\bf D46} (1992) 3062.
\medskip
\item {[38]} P.Colangelo, G.Nardulli, A.A.Ovchinnikov and N.Paver,
Phys. Lett. {\bf B269} (1991) 201.
\item{[39]} K.Berkelmann and S.Stone, report CLNS-91-1044 (1991).
\medskip
\item {[40]} M.E.Luke, Phys. Lett. {\bf B252} (1990) 447;
C.Glenn Boyd and D.E.Brahm, Phys. Lett. {\bf B257} (1991) 393.
\item{[41]} M.Neubert, report SLAC-PUB-5826 (1992).
\medskip
\vfill\eject\bye